\documentclass[pra,twocolumn,floatfix,superscriptaddress]{revtex4}
\usepackage{times,amsmath,amsfonts,amssymb,latexsym}

\usepackage{graphicx,epsf}
\usepackage{bm}
\usepackage{dsfont}
\usepackage{bbold}

\newcommand{\bra}[1]{\langle #1 |}

\newcommand{\ket}[1]{| #1 \rangle}

\newcommand{\norm}[1]{||#1||}

\begin{document}
\title{Excitonic diffusion length in complex quantum systems: The effects of
disorder and environmental fluctuations on symmetry-enhanced supertransfer}
\author{D. F. Abasto}
\affiliation{Department of Physics and Astronomy and Center for Quantum Information Science \& Technology, University of Southern California, Los Angeles, California 90089-0484, USA}
\author{M. Mohseni}
\affiliation{Center for Excitonics, Research Laboratory of Electronics, Massachusetts Institute of Technology, Cambridge, MA 02139, USA}
\affiliation{Disruptive Information Processing Technologies group, Raytheon BBN Technologies. 10 Moulton Street, Cambridge, MA 02138, USA}
\author{S. Lloyd}
\affiliation{Department of Mechanical Engineering, Massachusetts Institute of Technology, Cambridge, MA 02139, USA}
\author{P. Zanardi}
\affiliation{Department of Physics and Astronomy and Center for Quantum Information Science \& Technology, University of Southern California, Los Angeles, California 90089-0484, USA}

\date{May 20, 2011}

\begin{abstract}
Symmetric couplings among aggregates of $n$ chromophores increase
the transfer rate of excitons by a factor $n^2$, a quantum mechanical phenomenon called
``supertransfer.'' In this work we demonstrate how supertransfer effects induced by geometrical symmetries can enhance
the exciton diffusion length by a factor $n$  along cylindrically symmetric
structures, consisting of arrays of rings of chromophores, and along
spiral arrays.  We analyse both closed system dynamics and open
quantum dynamics, modelled by combining a random bosonic bath with
static disorder. In the closed system case, we use the
symmetries of the system within a short-time approximation to obtain a closed analytical expression
for the diffusion length that explicitly reveals the supertransfer
contribution.  When subject to disorder, we show that supertransfer can enhance excitonic
diffusion lengths for small disorders and characterize the crossover from coherent to incoherent motion. Owing to the quasi-1D nature of
the model, disorder ultimately localizes the excitons, diminishing but not destroying the effects of supertransfer.
When dephasing effects are included, we study the scaling of diffusion with both time and number of chromophores and observe that the transition from a
coherent, ballistic regime to an incoherent, random-walk regime occurs at the same point as the change from supertransfer to classical scaling.
\end{abstract}

\maketitle

\section{Introduction}

In 1954 Dicke introduced the phenomenon of \textit{superradiance},
a quantum interference effect induced by symmetries of spin-boson
interactions, in which many interacting atoms could collectively
conspire to yield an enhanced relaxation rate \cite{Dicke54}. Due to
this fundamental non-classical cooperation, the probability of a
single photon emission from a $n$ identical atoms collectively
interacting with vacuum fluctuations becomes n times larger than
incoherent individual spontaneous emission probabilities
\cite{RehlerEberly71}. Superradiant radiative relaxation can also
occur in molecular aggregates due to inherent coherent feature of
Frenkel exciton dynamics. This phenomenon can be observed when a
closely packed group of molecules interacting under certain symmetry
can collectively donate an excitation with a rate which is
much faster than each individual molecule. The adversarial effects
of inhomogeneous broadening and exciton-phonon interactions on such
cooperative relaxation in molecular systems have been studied in
detail \cite{Fidder91,Zhao99,Palacios02,Jin03}. 

The same symmetry principles that underlie superradiance can give
rise an analogous phenomenon known as
cooperative energy transfer or \textit{supertransfer} 
\cite{S,Scholes02,LM}. Generally, the exciton transfer rate can be calculated
from the transition probability of an excitation jumping from one
molecule to another using Fluorescence Resonance Energy Transfer (FRET), 
based on a 
perturbation treatment of the dipole-dipole interaction between individual
molecules. However, under strong and symmetrized interactions of a group of $n$ molecules the
excitation becomes highly delocalized, leading to
a large (effective) dipole moment associated with the entire group.
The resulting enhanced oscillator strength can lead to 
supertransfer when similar 
molecular assemblies,  with comparable effective dipole moments, exist 
that can play the role of acceptors. Under 
such conditions the rate of exciton dynamics should be 
calculated from these effective large dipole-dipole 
interactions to describe the coherent donation and 
acceptance among such molecular aggregates, with up 
to $n^2$ enhancement over the single molecule to single molecule
transfer rate, even in the far field \cite{LM}. 
A primary goal of this paper is to study the behaviour of superradiance, 
supertransfer, and other non-classical collective phenomena 
in presence of disorder and environmental fluctuations similar 
to those natural conditions of photosynthetic light-harvesting 
complexes \cite{Fidder91,Zhao99,Palacios02,Scholes02,Jang04}. 
Recently, the existence and role of quantum coherence in the 
dynamics of excitation energy transfer in biological systems 
have been of significant interest both experimentally 
\cite{Engel07,Lee07,Calhoun09,Mercer09,Scholes09-1,Scholes09-2,panit10} and theoretically \cite{mohseni-fmo,Rebentrost08-2,Olaya-Castro08,Rebentrost08-1,Plenio09,AkiPNAS,CaoSilbey,Caruso10,Sarovar,Fassioli10,Joel2010,Shabani2011,Mohseni2011,Shim2011}. 
Such studies can provide novel concepts and techniques 
for a deeper understanding of natural/engineering excitonic 
systems potentially leading to efficient and robust artificial 
light-harvesting \cite{Shabani2011,Mohseni2011}. 

A major problem in design and fabrication of novel excitonic devices is the
limited exciton diffusion length that could be of about 10 nm in disordered
materials. This issue has lead to low efficiency and complicated device
structures in organic photovoltaic cells \cite{pemans03,lunt09}, and it is a bottleneck in the
performance of excitonic transistors \cite{high08} and organic light emitting diodes \cite{castellano10}.
A key open question is whether one can use 
quantum-mechanical supertransfer effects to enhance exciton 
diffusion length in such disordered systems.
Recent experimental investigations of nano-engineered 
biological systems suggest that under laboratory conditions 
certain aspects of photosynthetic
complexes can be emulated that could be potentially exploited 
for efficient energy transport \cite{E,MPF}. In particular, 
excitonic diffusion lengths up to a micron have been recently reported in
engineered arrays of LH2 complexes \cite{E}. Self assembled ring
structures containing fluorescent chromophores attached to tobacco
mosaic virus coat monomers exhibit efficient exciton transport together
with a broad spectrum light collection
with over 90\% efficiency \cite{MPF}.
It is of considerable interest to explore whether the energy transport 
mechanism in such systems was facilitated in part due to
symmetries in the arrangement of chromophores, which give could rise to
the phenomenon of supertransfer \cite{S,Scholes02,LM}. 

In this paper, we investigate excitonic transport
in systems consisting of rings of chromophores stacked in cylindrical arrays,
as a function of the number of chromophores per ring, the spacing
between rings, and the strength of decoherence and disorder.
We also investigate excitonic transport in dipole-coupled spiral
structures.   Such geometries are relevant not only to the
experimentally investigated systems mentioned above, but to
naturally occurring cylindrical arrays of chromophores such as the
green sulphur bacterium chlorosome.  We use the symmetries of
the system to derive analytic solutions for the behavior of
the closed system in the absence of environmental interactions,
and perform simulations to capture the dynamics of excitonic
diffusion in the presence of environmentally-induced noise and
disorder.  Our results provide clear evidence for the presence
of supertransfer in the appropriate regimes and for the destruction
of supertransfer in other regimes.

Specifically,  we study the effect of supertransfer on
the diffusion length $\sigma$ of an initially delocalized exciton
along a linear chain of chromophoric rings and helical rods, mediated by dipolar couplings
between chromophores.   Supertransfer-induced
enhancements in the hopping rate translate into a
commensurate increase in the distance travelled by the excitons.
The strength of the supertransfer effect depends on
the number of chromophores per ring $n$ -- higher $n$ yields
higher supertransfer rates,  and on the distances
between rings --  smaller inter-ring distances yields larger
asymmetries and diminishes supertransfer.
By changing the strength of the interaction with the environment,
we investigate excitonic transport in both coherent/ballistic and
incoherent/diffusive regimes.  By changing the degree of disorder,
we investigate how the degree of localization depends on $n$.
The analysis is carried out analytically in the closed system case,
and numerically in the presence of disorder and interactions with
a bosonic bath.  Our analytic and numeric solutions allow us
to investigate scenarios where
rings are packed closely in a cylinder, and where chromophores
are arranged in spirals.
\section{Theoretical Model}
\subsection{A simple example}
The concept of supertransfer can be easily reviewed in the case of two spin systems $A$ and $B$, each containing $n_A$ and $n_B$ sites respectively, invariant under permutation symmetry \cite{LM}.   In the presence of such symmetric couplings, the
hopping rate from the symmetrized single-excitation state of $A$
to the symmetrized single-excitation state
of $B$ is $n_A n_B$ times the hopping rate of a
localized excitation from one of the $A$ to one of the $B$ sites.
(The effect can be even larger -- up to $n_A^2 n_B$
-- for multiple excitation
states \cite{LM}.  For simplicity, this paper
will focus on single exciton states.)
Supertransfer can be captured in a simple way
by the following symmetric hopping Hamiltonian:
\begin{equation}\label{SimHam}
H=-\frac{\epsilon_A}{2}\sum_{i=1}^{n_A}\sigma_z^i - \frac{\epsilon_B}{2}\sum_{j=1}^{n_B}\sigma_z^j + \gamma\sum_{i=1,j=1}^{n_A,n_B}\sigma^i_+\sigma^j_-+\sigma^i_-\sigma^j_+.
\end{equation}
Here, $i$ labels the $n_A$ sites, and $j$ the $n_B$ sites of $A$ and $B$,
respectively. Introducing the angular momentum operators
\begin{equation}
J^3_A=\frac{1}{2}\sum_i^{n_A}\sigma_z^i, \quad J^{\pm}_A=\sum_i^{n_A}\sigma^i_{\pm}\label{Ja},\quad J^{\pm}_B=\sum_i^{n_B}\sigma^i_{\pm},
\end{equation}
the previous Hamiltonian can be expressed as
\begin{equation}
H=-\epsilon_A J^3_A - \epsilon_B J^3_B + \gamma(J^+_AJ^-_B + J^-_AJ^+_B)
\end{equation}
From the above expressions, the probability amplitude for a single
symmetrized excitation state $\ket{\phi_A}$ over the system $A$,
$\ket{\phi_A} = \frac{1}{\sqrt{n_A}}
\sum_{i=1}^{n_A}|i\rangle$,
to hop to a corresponding symmetrized state over system $B$
is given by $\gamma\sqrt{n_An_B}$, in first order perturbation theory.
The corresponding transition probability $\gamma^2n_An_B$ is
therefore $n_An_B$ times the probability $\gamma^2$ for a
localized exciton state $\ket{1}$ to hop from one of the $A$
sites to any one of the $B$ sites. This enhancement effect
in the hopping rate is called supertransfer \cite{S,Scholes02,LM}.   The key feature
of supertransfer is coherence within the individual systems $A$
and $B$ respectively.  Supertransfer between $A$ and $B$ can either be
coherent or incoherent, depending on the spatial separation of systems and strength of the interaction
with the environment.

\subsection{Exciton supertransfer in cylindrical geometries}
Here, we study supertransfer effects in cylindrical
aggregates of sites interacting via a $1/r^3$ potential.
We apply our analysis to the transition
dipoles of an array of chromophores to look at excitonic hopping
through the array.
We consider diffusion in the context of two types of geometrical set-ups. In the first configuration, the sites are arranged in rings, which are stacked in a cylinder.
Each ring has a radius $R$, consists of $n$ dipoles each.
There are a total of $N$ rings, arranged co-axially in a cylinder
with a separation $D$ between adjacent rings.
The position of the $i$'th site inside the $j$'th ring
is given by $r_{i}=(Rcos(2\pi i/n),Rsin(2\pi i/n), jD)$, for $i=1,2, \dots, n$ and $j=1,2, \dots, N$. In the second geometry, chromophores form helical rods, with their positions given by $r_{i}=(Rcos(2\pi i/n),Rsin(2\pi i/n), di/n)$, where $d$ is the pitch of the helix, $R$ its radius, $n$ the number of chromophores per turn, with $N$ turns in total, and $i=1, 2, \dots, nN$.

The single exciton manifold approximation will be employed,
so that the system Hilbert space is $\mathcal{H}=\mathbb{C}^{nN}$,
spanned by a basis $\mathcal{S}=\{|m\rangle\}_{m=1}^{nN}$,
where $\ket{m}$ denotes the state in which the $m^{th}$ molecule is
excited.

The Hamiltonian $H_T$ of our ring aggregates is of the form
\begin{equation}\label{HamSym}
H=\sum_{r}(\langle E \rangle + \eta_r)|r\rangle \langle r| + \sum_{m\ge r}J_{mr}(\ket{r}\bra{m}+\ket{m}\bra{r}),
\end{equation}
with all the interactions included.
For simplicity, we initially consider an interaction of the
form $J_{mn} = J / r_{\mathrm{nm}}^3$,
with $r_{mn}=|\vec{r}_{m} - \vec{r}_{n}|$. In what follows, the normalization $J=1$ will be
adopted, with time measured in units of $1/J$.
This interaction is a simplified version of the dipolar interaction
that has the same distance dependence as the conventional dipolar
interaction, but that does not take into account the orientations
of the dipoles. This simplified assumption will be relaxed
below.

The influence of the environment over the system will be considered
by including both on-site energy disorder and on-site dephasing (Haken-Strobl
model) which has recently been widely used for studying environment effects on energy transport in light-harvesting systems, e.g., see Refs  \cite{Rebentrost08-2,Plenio09,CaoSilbey}.  In (\ref{HamSym}), $\langle E \rangle$ represents the average
molecular excitation energy, which can be dropped by shifting
all the energies by $\langle E \rangle$. The static inhomogeneous
offset $\eta_r$ in the energy of the $r$'th site
reflects the disorder caused by the surroundings. We take
$\eta_r$ to be a Gaussian random variable with standard
deviation $\Sigma$ and probability distribution given by
\begin{equation}\label{disor}
P(\eta_r) = \frac{1}{\Sigma\sqrt{2\pi}}\exp\big(-\eta_r^2/2\Sigma^2\big),
\end{equation}
where we ignore correlations between the offsets of each molecule.
The strength of the disorder $\Sigma$ is measured in units of $J$.

The effects of the bosonic bath surrounding the system are
be modelled by an on-site dephasing model or Haken-Strobl model,
given by a Lindblad superoperator of the form
\begin{eqnarray}
\mathcal{L}_{\mathrm{deph}}\rho(t)=\gamma\sum_{n=1}\big[S_n\rho(t)S_n-
S_n\rho(t)/2 - \rho(t) S_n /2\big]
\end{eqnarray}
with the sum running over all the sites, $S_n=\ket{n}\bra{n}$
and $\gamma$ the dephasing rate. Thus, the total dynamics is given by
\begin{equation}\label{mastereq}
\frac{d \rho(t)}{dt} = -\frac{i}{\hbar}[H ,\rho(t)] + \mathcal{L}_{\mathrm{deph}}(\rho(t)) - \{H_{\mathrm{recom}},\rho(t)\},
\end{equation}
where the effects of exciton recombination is captured by the
final term.  Here, $H_{\mathrm{recom}}=\kappa\sum_n\ket{n}\bra{n}$,
with $1/\kappa$ the lifetime of the exciton.
\begin{figure*}[t]
\includegraphics[scale=0.18]{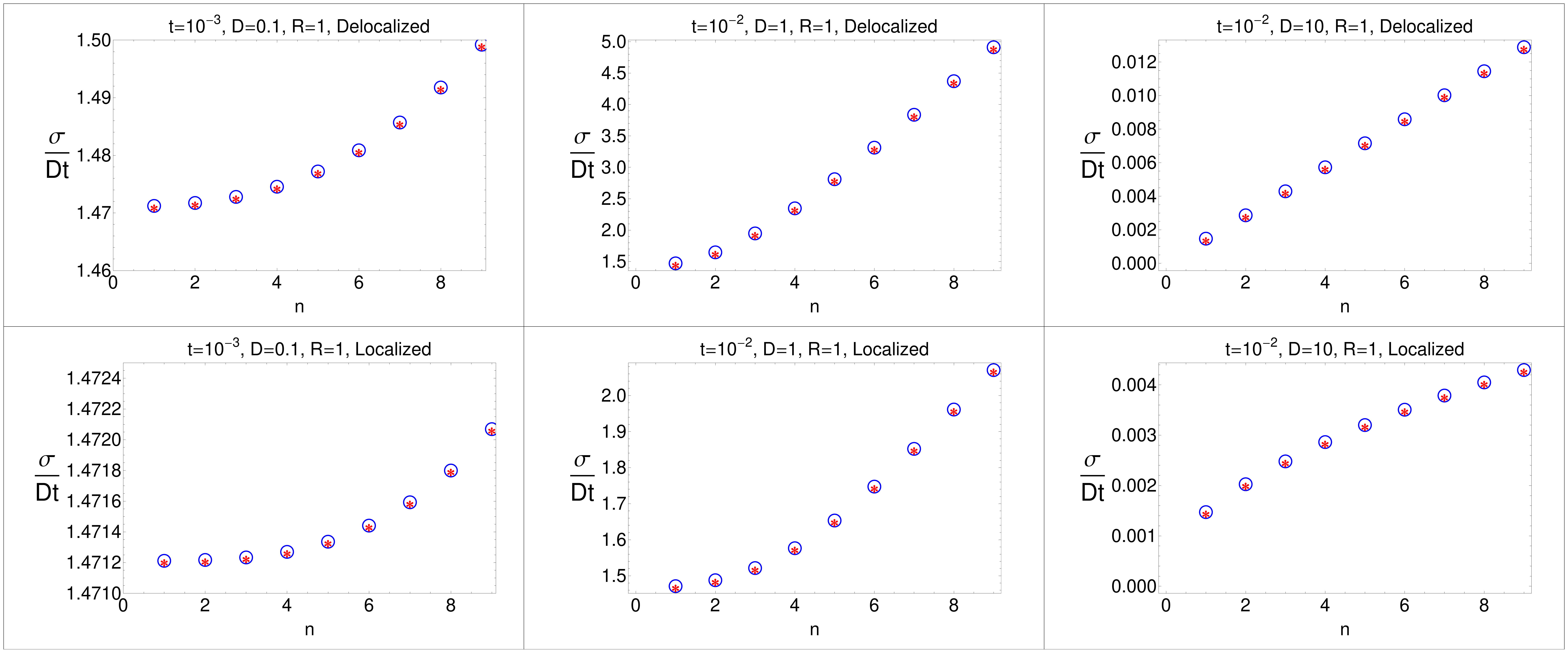}
\caption{(color online) Exciton diffusion in a chain of stacked rings as a function of number of nodes per ring $n$, per unit time and distance $D$ between adjacent rings. A comparison is made between the diffusion obtained numerically (blue circles), and calculated via the theoretical formulas (\ref{sigmadeloc}) and (\ref{sigmaloc}) (red stars).  The upper (lower) panel shows diffusion obtained for an initial delocalized (localized) state with support over the middle ring of the chain. When the exciton is initially delocalized and the rings are far apart from each other as compared to their radius (the far-field regime $D/R=10$), the set of inter-ring couplings are symmetric under permutations of the chromophores, and therefore diffusion grows linearly with the number of chromophores per ring $n$, as expected from supertransfer arguments. Meanwhile, if the exciton is initially localized over a single chromophore on the middle ring of the chain, the diffusion grows with the classical $\sqrt{n}$ scaling. When the rings are close to each other as compared to their radius (the near-field regimes $D/R=0.1$ and $D/R=1$), the diffusion has an anomalous scaling with the number of sites n, i.e. $\sigma\sim n^\alpha$, $\alpha>1$, for both types of initial states. This is a finite size effect, as explain in the main text. We employed $N=31$ rings. In all the cases, there is a good agreement between theory and numerical simulations. Note that for $D/R=0.1$, both the time and coupling strength were adjusted in order to avoid boundary effects.}
\label{Predictions}
\end{figure*}
\begin{figure*}[t]
\includegraphics[scale=0.18]{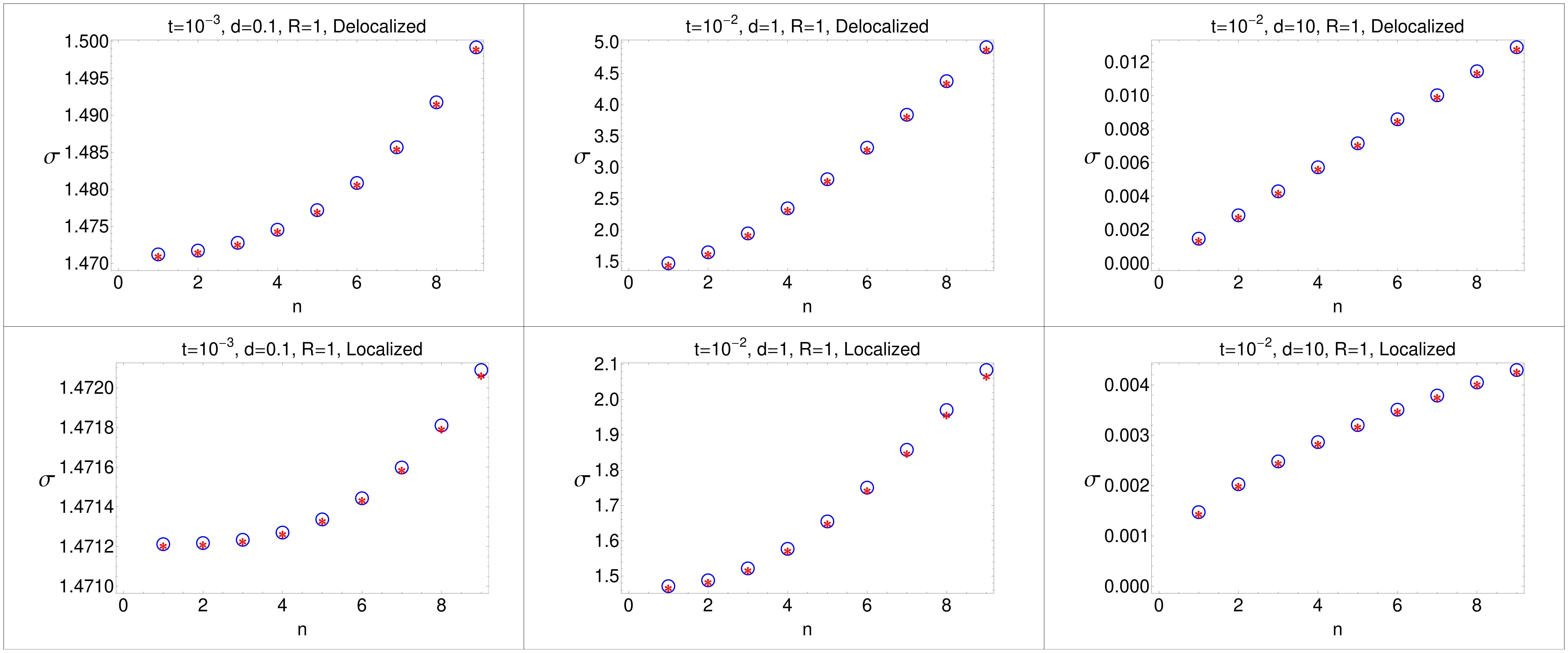}
\caption{(color online) Exciton diffusion along a spiral chain of chromophores as a function of number of nodes per ring $n$, per unit time and pitch or aperture $d$. The spiral structure is approximated by a stacked ring geometry, in which each turn of the helix is replaced by a ring, with as many chromophores per ring as nodes per turn in the helix, and distance $D$ between rings equal to the pitch $d$ of the helices. The numerical results of diffusion obtained from solving eq.(\ref{mastereq}) (blue circles) for the helical rod is compared with the diffusion along the corresponding approximating circular structure, using formulas (\ref{sigmadeloc}) and (\ref{sigmaloc}) (red stars). The upper panel shows diffusion obtained for an initial delocalized state over a set $n$ of contiguous chromophores along a full turn of a helix, while in the lower panel an exciton with support over a single chromophore was employed.
When the exciton is initially delocalized and the pitch of the helix is bigger than the radius (the far-field regime $d/R=10$), the diffusion grows linearly with the number of chromophores per ring $n$ as expected from supertransfer arguments. Meanwhile, if the exciton is initially localized over a single chromophore on the middle ring of the chain, the diffusion grows with the classical $\sqrt{n}$ scaling. When the rings are close to each other as compared to their radius (the near-field regimes $d/R=0.1$ and $d/R=1$), the diffusion has an anomalous scaling with the number of sites n, i.e. $\sigma\sim n^\alpha$, $\alpha>1$, for both initial states. This is a finite size effect, as explain in the main text. The approximation of the spiral structure via a set of facing rings works very well, specially for the delocalized case. We employed $N=31$ turns/rings in our numerical simulation. Note that for $D/R=0.1$, both the time and coupling strength were adjusted in order to avoid boundary effects. The values of diffusion are given per unit time and pitch $d$.}
\label{RingsApprox}
\end{figure*}
\begin{figure}[t]
\includegraphics[scale=0.38]{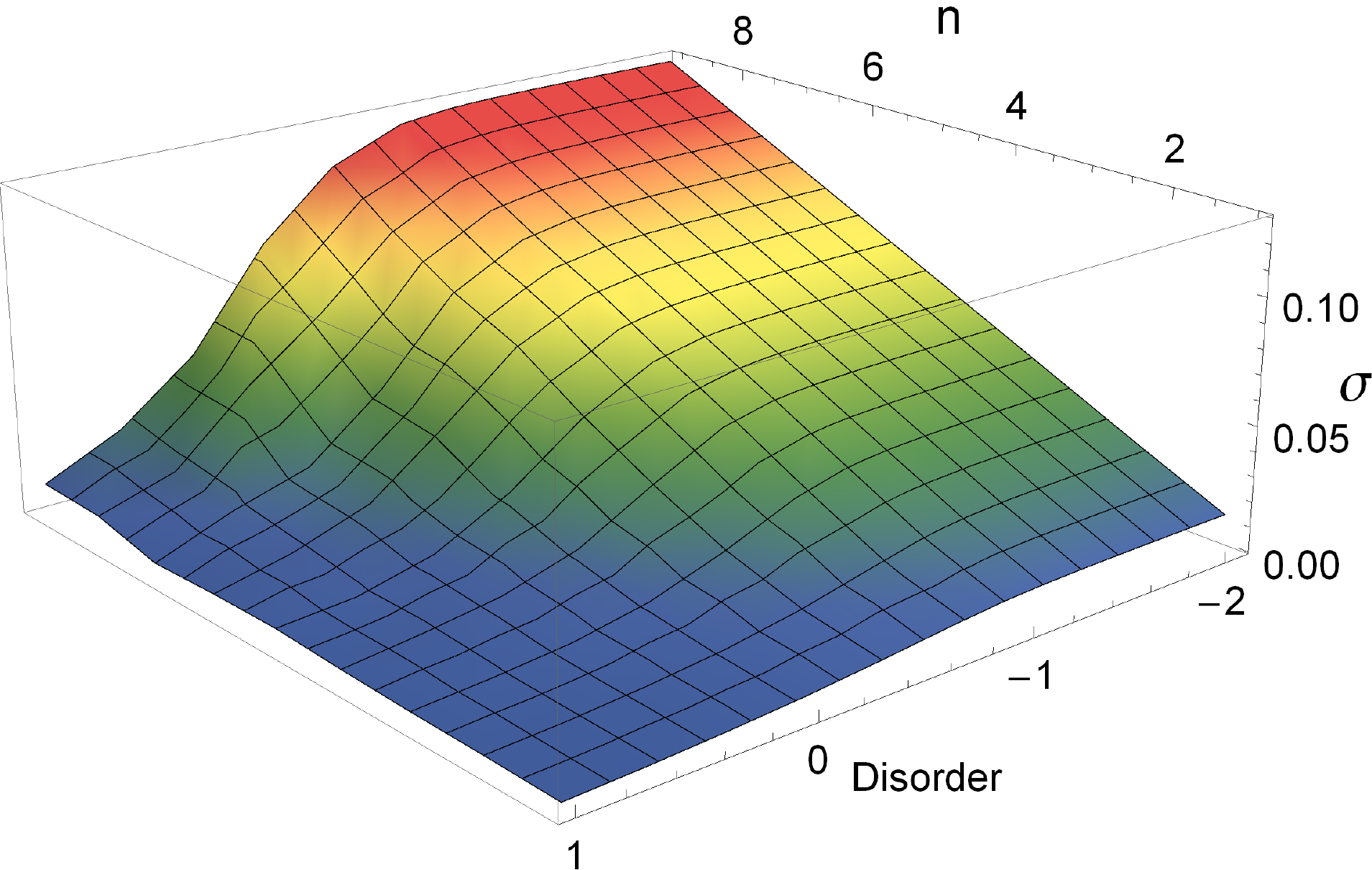}
\caption{(color online) Diffusion length $\sigma$, along a system of stacked rings arranged co-axially in a cylinder, as a function of the number of chromophores $n$ per ring and on-site energy disorder $\Sigma$ that is represented in a logarithmic scale in units of $J=1$. The system parameters are $(N,R,D)=(31,1,10)$. Each point corresponds to 500 realizations. The time was set at $t=1$. For small times or disorder, $t\Sigma\ll 1$, we observe a linear scaling of diffusion with respect to the number of chromophores $n$, a clear signature of supertransfer, as the system evolves coherently. However, for long times or large on-site energy disorder, $t\Sigma\gg 1$, diffusive behaviour sets in and the mismatch among the on-site energies causes localization, regardless of the number of nodes per ring $n$.}
\label{DiffVsnVsDisor}
\end{figure}

\begin{figure}[t]
\includegraphics[scale=0.25]{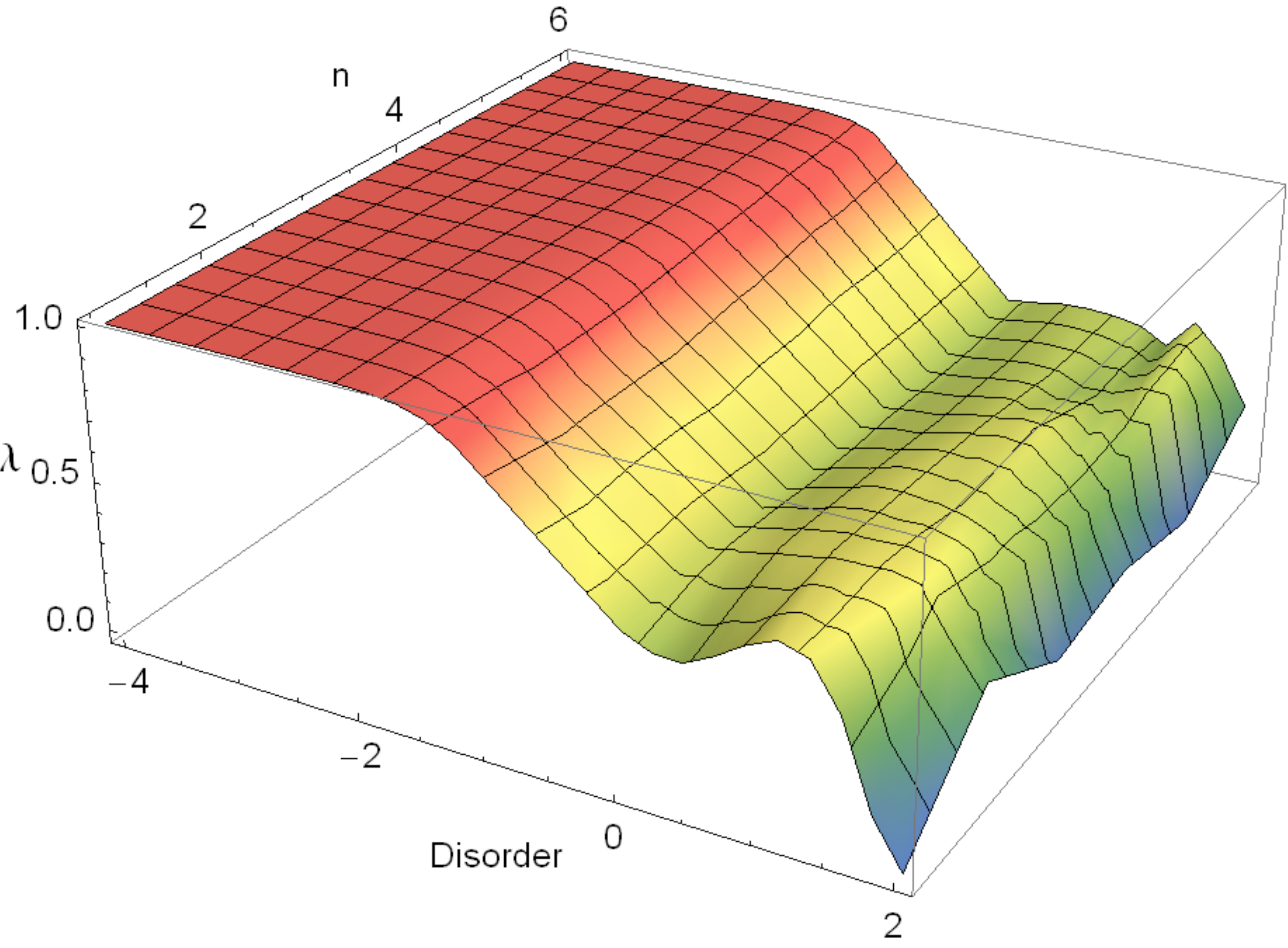}
\caption{(color online) Scaling exponent $\lambda$ of diffusion with time, $\sigma(t)\propto t^{\lambda}$, along a system of stacked rings arranged co-axially in a cylinder, as a function of the number of chromophores $n$ per ring and on-site energy disorder $\Sigma$. The disorder is represented in a logarithmic scale in units of $J=1$ and the time was set at $t=1$. The parameters are $(N,R,D)=(31,1,10 )$. The exponent $\lambda$ signals the character of the exciton propagation, with $\lambda=1$ representing quantum coherent or ballistic spreading, while $\lambda=1/2$ signaling classical or diffusive evolution. We observe that the exciton motion is coherent for small values of disorder, and shows a crossover from coherent to incoherent around $\Sigma\sim 10^{-2}$, independently of the number of nodes per ring $n$. For disorder $\Sigma\gg 1/t$ the diffusion shows fluctuations around the classical random walk value $\lambda=1/2$.}
\label{LambdaInDiffVsnVsDisor}
\end{figure}
\begin{figure}[th]
\includegraphics[scale=0.33]{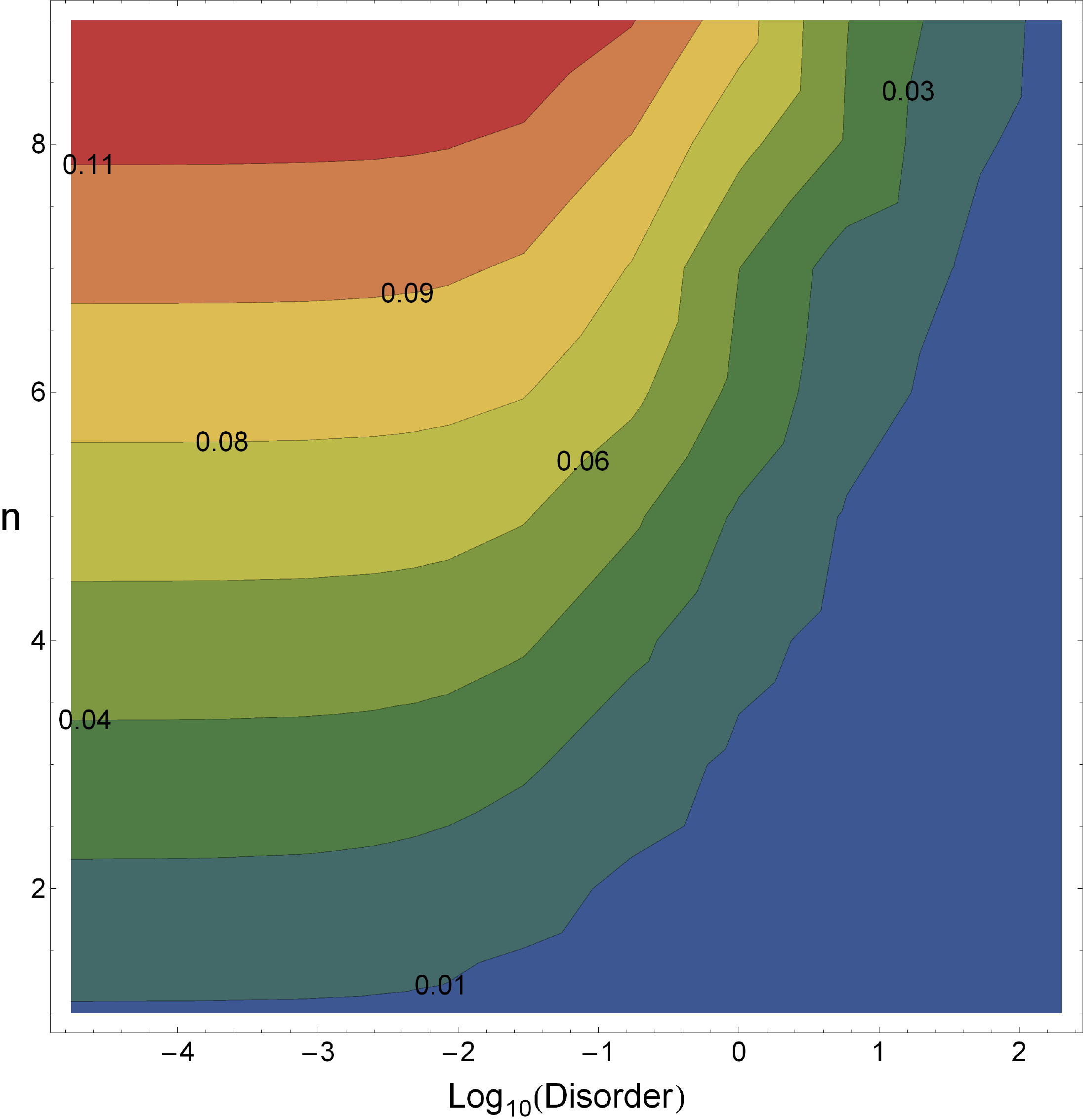}
\caption{(color online) Level curve for diffusion $\sigma$ along a chain of stacked rings, as a function of disorder $\Sigma$ and number of chromophores per ring $n$. This figure was obtained from Fig. (\ref{DiffVsnVsDisor}). For on-site energy disorder $\Sigma< 10^{-2}$, the exciton still evolves coherently and its diffusion is independent of disorder, with a supertransfer scaling linear in $n$. Around $\Sigma\sim10^{-2}$ we have a crossover from ballistic to diffusive behaviour, and a $\sqrt{n}$ type of behaviour sets in. This crossover agrees with Fig. (\ref{LambdaInDiffVsnVsDisor}). In the classical regine, for a fixed value of diffusion, the system is more resilient to disorder as n increases.}
\label{LevelCurveDisor}
\end{figure}
\begin{figure}[t]
\includegraphics[scale=0.47]{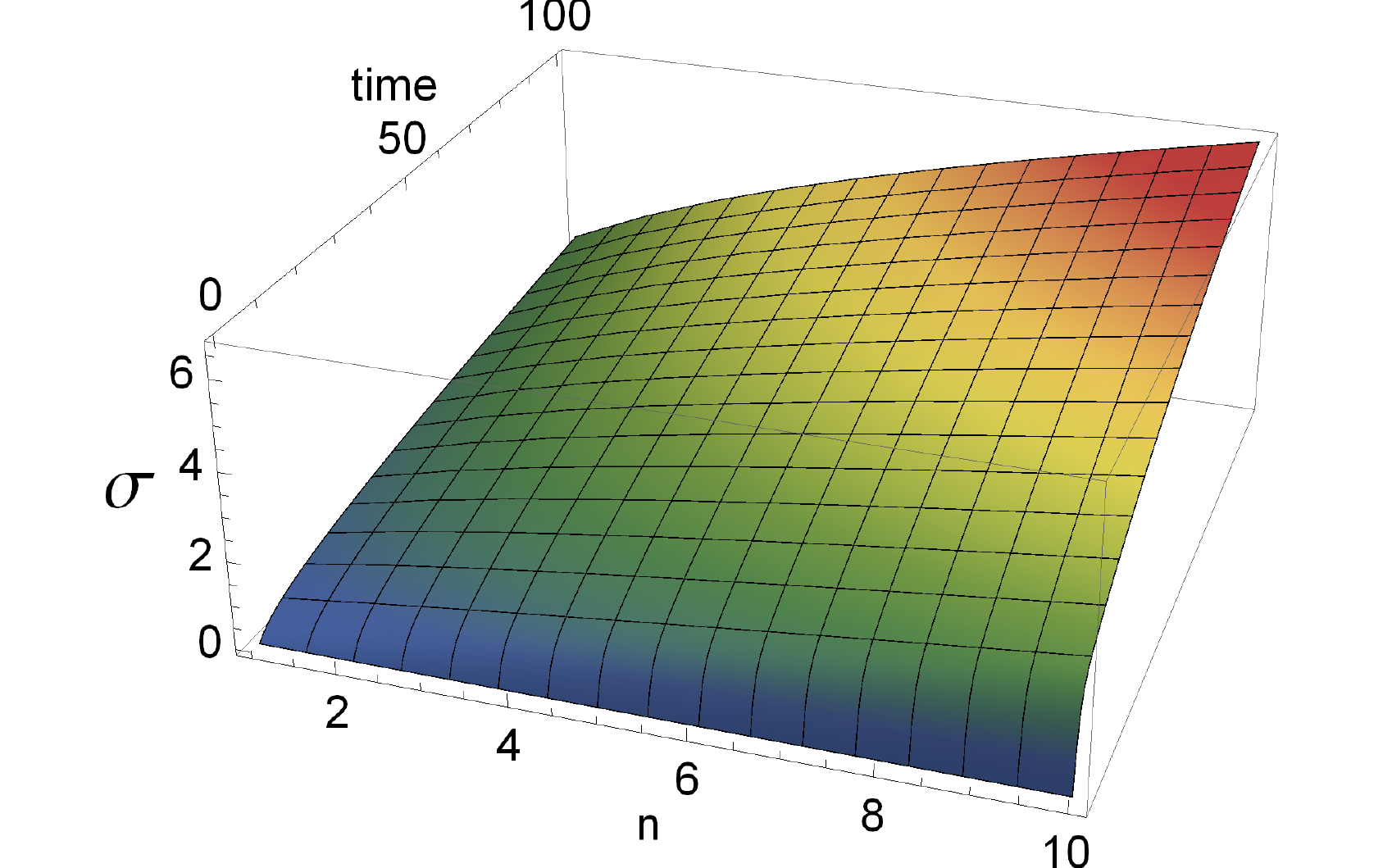}
\caption{(color online) Diffusion length along a chain of stacked rings, as a function of number of nodes per ring $n$ and time, in the presence of on-site dephasing. The dephasing strength $\gamma$ was set at $\gamma=1$. An initial delocalized state was employed for the numerical simulations. We observe that for times short compared to the dephasing time, $t<1/\gamma$, the motion is still coherent and the diffusion is proportional to the number of chromophores per ring $n$, $\sigma(t,n)\sim n$, a clear signal of supertransfer effects. Meanwhile, for long times or strong dephasing,  $t>1/\gamma$, the quantum coherence is destroyed, with diffusion exhibiting the classical scaling $\sigma(t,n)\sim \sqrt{n}$. The parameters used are $(N,R,D)=(31,1,10)$. The axis for $\sigma$ was multiplied by a factor of 100.}
\label{svstg}
\end{figure}
\begin{figure*}[t]
\includegraphics[scale=0.18]{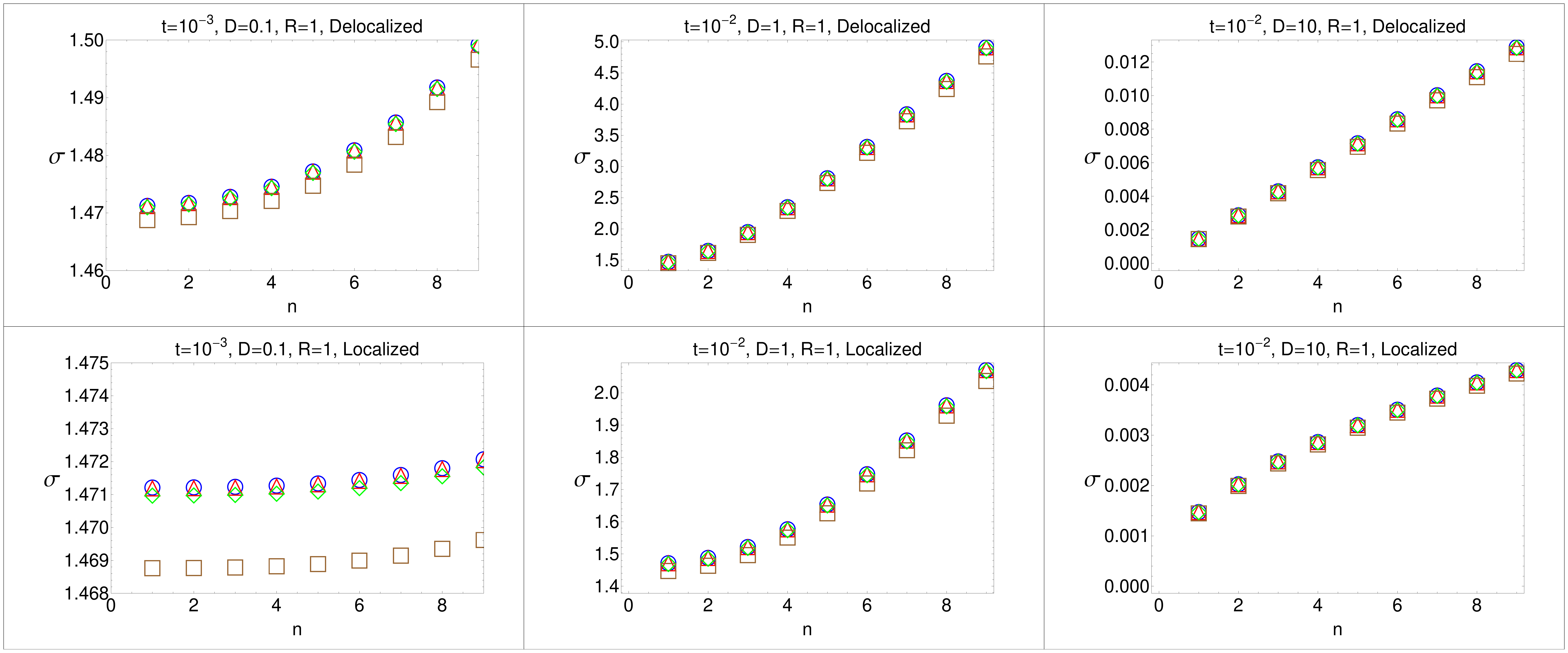}
\caption{(color online) Effects of dephasing on diffusion. Exciton diffusion in a chain of stacked rings as a function of number of nodes per ring $n$, per unit time and distance $D$ between adjacent rings. We compare the theoretical predictions for diffusion (\ref{sigmadeloc}) and (\ref{sigmaloc}) (blue squares) with numerical simulations under different values of on-site dephasing: $\gamma=0.1$ (red triangles), $\gamma=1$ (green diamonds) and $\gamma=10$ (brown squares). We employed $N=31$ rings and time was fixed at $t=1$. The upper (lower) panel shows diffusion obtained for an initial delocalized (localized) state with support over the middle ring of the chain. When dephasing is included, it does not modify the theoretical predictions eq. (\ref{sigmadeloc}) and (\ref{sigmaloc}), up to second order in the expasion of $e^{t\mathcal{L}}\rho(0)$ in time. Therefore, for values of dephasing small compared with time ($\gamma=0.1$) the agreement between the numerics and analytics is
 quite good.}
\label{PredVsDeph}
\end{figure*}

\begin{figure}[t]
\includegraphics[scale=0.27]{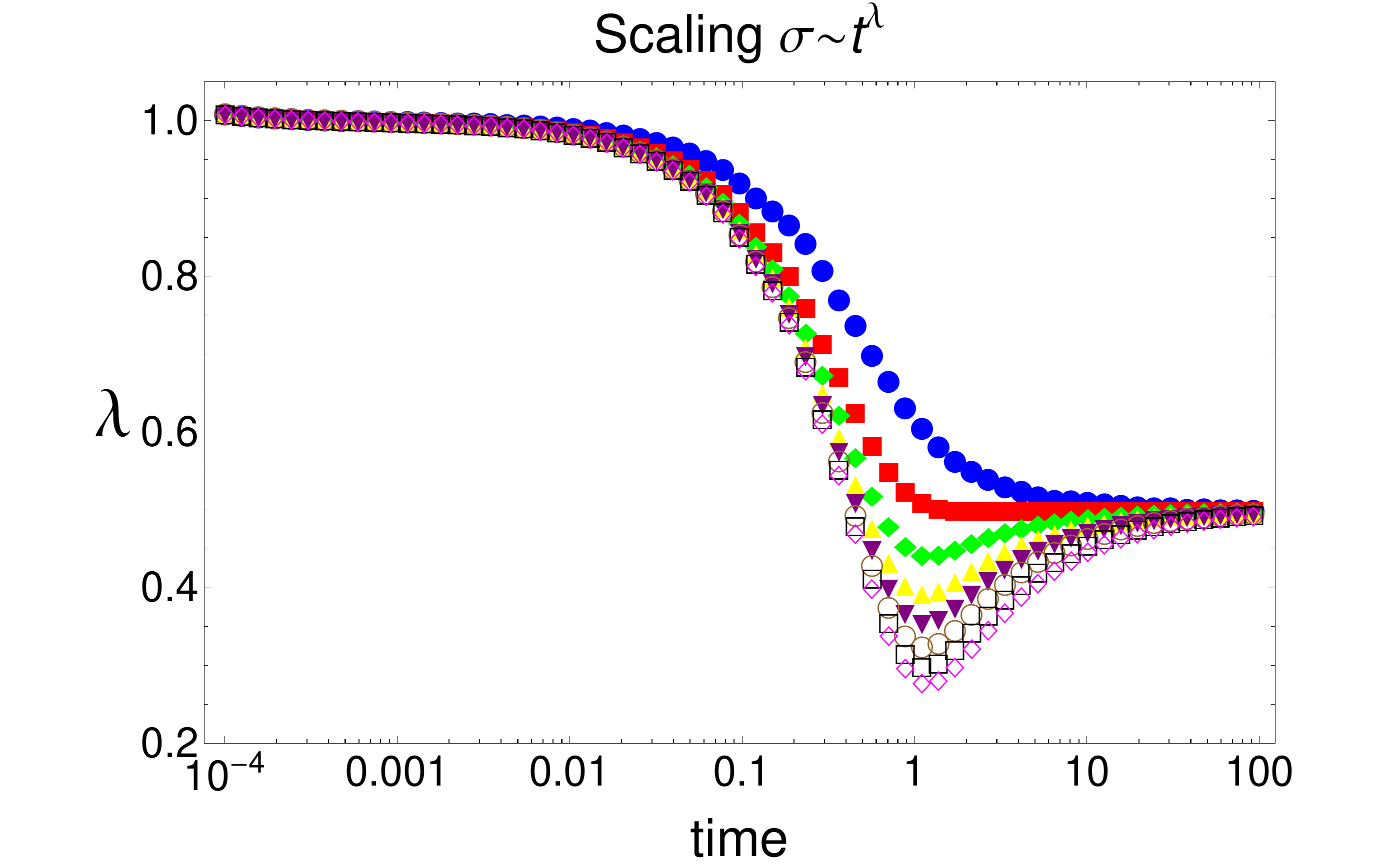}
\caption{(color online) Scaling exponent $\lambda$ of diffusion with time, $\sigma(t)\propto t^{\lambda}$, for an initial delocalized exciton moving along a system of stacked rings, as a function of time, for $n=1$ (blue circles), to $n=6$ (pink diamonds). On-site dephasing was included, with parameter $\gamma=5$. The exponent $\lambda$ signals the character of the exciton propagation, with $\lambda=1$ representing quantum coherent or ballistic spreading, while $\lambda=1/2$ signaling classical or diffusive evolution. We observe the exciton exhibits a ballistic spreading $\sigma(t)\propto t$ ($\lambda=1$) at small times $t\gamma\ll 1$, converging towards a classical random walk behaviour $\sigma(t)\propto \sqrt{t}$ ($\lambda=1/2$) as $t\gamma \gg 1$, with the crossover around $t\sim 1/\gamma$. We lack an explanation for the dip of the scaling exponent $\lambda$ below 0.5 around $t=1$, which may point to an intriguing physical feature, and is left for future research. The ring
 parameters were set to $(N,R,D)=(31,1,10)$.}
\label{LambdaVsTime}
\end{figure}

\begin{figure}[t]
\includegraphics[scale=0.27]{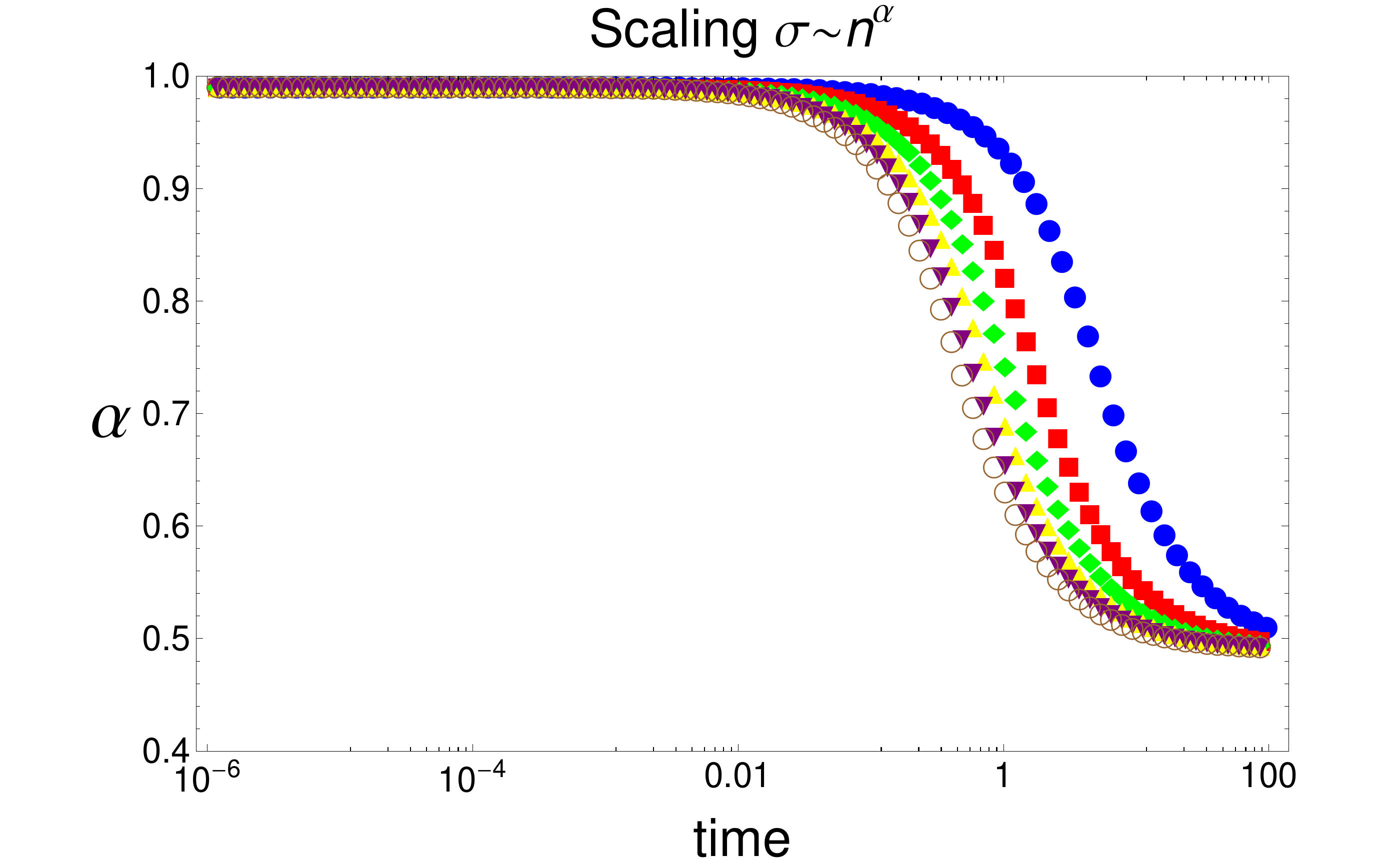}
\caption{(color online) Scaling exponent $\alpha$ of diffusion with number of chromophores $n$, $\sigma(n)\propto n^{\alpha}$, for an initial delocalized exciton moving along a system of stacked rings, as a function of time, for on-site dephasing strength $\gamma=1$ (blue circles) up to $\gamma=11$ (open brown circles). The exponent $\alpha$ provides an indication of supertransfer efffects when $\alpha=1$, while $\alpha=1/2$ signals classical or diffusive evolution. We observe the exciton exhibits a supertansfer regime $\sigma(n)\propto n$ ($\alpha=1$) at small times $t\gamma\ll 1$, converging towards a classical behaviour $\sigma(n)\propto \sqrt{n}$ ($\alpha=1/2$) as $t\gamma \gg 1$. The crossover from one regime to another occurs around $t\sim 1/\gamma$, coincident with the transition from ballistic to diffusive regime in Figure (\ref{LambdaVsTime}). The ring parameters were set to $(N,R,D)=(31,1,10)$, and we employed $n=1, 2, \dots, 10$ nodes per ring in order to perform the fitting $\sigma(n)\sim n^{\alpha}$.}
\label{AlphaVsTime}
\end{figure}

\subsection{Choice of initial state and diffusion length}
By its definition, supertransfer relies fundamentally on two different symmetries: the permutation symmetry
present on the Hamiltonian, as well as the permutation symmetry
and coherence on the initial quantum states. Both the ring chain and helical rods set-ups resembles the structures of the fluorescent
chromophores attached to tobacco mosaic virus monomers, which could self-assemble
into stacks of rings or helices \cite{MPF}. Note that the invariance
under permutations of the $n$ dipoles inside a ring resembles
the symmetric interaction in (\ref{SimHam}) for $D>R$, where
strong supertransfer effects should take place.
When the rings become closer, $D<R$, then the interaction
between rings becomes less symmetric.  The strength of
supertransfer is diminished, but the effect persists,
as we will show analytically below.

As the initial quantum state in the equations of motion, we choose
a delocalized quantum state over the ring of $n$ sites,
of the form $\rho_{\mathrm{Deloc}}=\ket{\phi}\bra{\phi}$,
with $\ket{\phi}=\frac{1}{\sqrt{n}}\sum_{i=1}^{n}\ket{i}$,
supported on the middle ($N/2$) ring of the $N$-ring chain.
We consider an odd number $N$ of rings, each with an
odd number of sites $n$, with $N=2T+1$, $n=2t+1$.
The rings are labelled by $r=-T,\dots, -1, 0, 1, \dots, T$,
with $0$ denoting the middle ring.  (The use of odd numbers
of rings and sites are simply for the convenience of labelling.)
To quantify how much an initial excitation located at the
middle ring diffuses towards the boundaries, we solve for
$\rho(t)$ in the master equation (\ref{mastereq}) and calculate
the second moment $\sigma(t)$ of the probability distribution $p_i(t)$ for an exciton to be at the $i^{th}$ ring at time $t$:
\begin{equation}\label{diff}
\sigma(t)=D\sqrt{\sum_{r=-T}^{T} p_{r}(t)r^2}
\end{equation}
with $p_r(t)=\sum_{j=1}^{n}\rho_{jj}(t)$ the probability
for the exciton to be present in the $r^{th}$ ring at time $t$,
$\rho_{jj}(t)=\bra{j}\rho(t)\ket{j}$ the exciton population of
site $j$ on ring $r$ and $D$ the distance between two adjacent rings.
The diffusion length $\sigma(t)$ measures how far the exciton diffuses
along the system of rings.
We begin by presenting our analytic model for the closed quantum
system to show that supertransfer enhances the diffusion of a delocalized initial state by a factor
$n$.  This enhancement for the closed system (coherent evolution and
ballistic transport) will be found to persist in the open
system, at least for short times.

\section{Closed system dynamics: Analytical calculation of diffusion}

We will proceed to calculate $\sigma(t)$ for a closed system
using a short time approximation, that is
$t||H||\ll 1$, with $||H||$ the operator norm of the Hamiltonian $H$ (maximum eigenvalue of $H$). We assume long lifetime
$1/\kappa  \gg t$, such that recombination effects can be ignored.
An important observation is that the Hamiltonian of the linear
chain of rings is very close to a block circulant structure.
A block circulant matrix B $\in \mathcal{BC}_{N,n}$ is of the from

\begin{equation}
B=\mathrm{circ}(\bf{b}_0 \, \bf{b}_1\, \dots \, \bf{b}_{N-1})=\begin{pmatrix} \bf{b}_0 & \bf{b}_1 & \dots & \bf{b}_{N-1} \\
                                                                              \bf{b}_{N-1} & \bf{b}_0 & \dots & \bf{b}_{N-2} \\
                                                                              \vdots & \vdots & \dots  & \vdots \\
                                                                              \bf{b}_{1} & \bf{b}_2 & \dots & \bf{b}_{0} \\ \end{pmatrix}
\end{equation}
where the $n\times n$ matrices $\bf{b}_i$ are themselves circulant,
that is $\mathbf{b_i}=\mathrm{circ}(b_{i,0}\, b_{i,1}\, \dots\, b_{i,n-1})$ \cite{D}.
In this context, $\bf{b}_0$ represents the Hamiltonian for a single ring,
while the matrices $\bf{b}_i$, $i\ne 0$ represents the $n \times n$ set
of interactions between two rings. The circularity of $\bf{b}_0$ is ensured
by the symmetry of the ring structure.  A symmetric, block circulant
matrix B has the property that $\bf{b}_i=\bf{b}_{N-i}$, $i=0, 1, \dots, N-1$.
This property would hold exactly if we were to impose periodic boundary
conditions along the chain, so that the system is topologically
a `ring of rings' or torus.  The actual
 chain Hamiltonian $H$, does not possess periodic boundary conditions.
Nevertheless, for times short enough so that boundary effects can be neglected, we will see that the analytical expression for diffusion $\sigma$ obtained assuming block-circularity matches quite well the numerical results.

A matrix B $\in \mathcal{BC}_{N,n}$ if and only if it commutes
with the unitary matrix $\Pi_{N,n}= \pi_N \otimes\mathbb{1}_n$, where $\pi_N$ is the fundamental $N\times N$ permutation matrix  $\pi_N=\mathrm{circ}(0\, 1\, \dots\, 0)$.

Let us then consider the matrix $H\in \mathcal{BC}_{N,n}$ as being block circulant, with circulant blocks. Such Hamiltonian will have the form $H=\mathrm{circ}(\bf{H}_0 \, \bf{H}_1\, \dots \, \bf{H}_{T}\, \bf{H}_{T} \, \dots \, \bf{H}_1)$, where ${\bf{H}_j}\equiv \mathrm{circ}(h_{j,k})_{k=1}^{n}$.
The spectrum of such matrices is well-known \cite{T}. Let $w_j=\exp(2\pi i j/N)$, and $\rho_k=\exp(2\pi i  k/n)$. Define the matrix $F_N(\omega)$ given by
\begin{equation}
F_N(\omega)=\begin{pmatrix} 1 & 1 & 1 & \dots & 1\\ 1 & w_1 & (w_1)^2 & \dots & (w_1)^{N-1} \\ . & . & . & \dots & . \\1 & w_{N-1} & (w_{N-1})^2 & \dots & (w_{N-1})^{N-1}\\\end{pmatrix}.
\end{equation}
Then, the circulant matrix $H$ can be diagonalized by $F_N(\omega)\otimes F_n(\rho)$, with eigenvalues $e(p,q)$ given by
\begin{eqnarray}\label{eigenvalues}
e(p,q)&=\sum_{j=0}^{N-1}\sum_{k=0}^{n-1}(w_j)^p (w_k)^q h_{j,k}\\
      &= \sum_{j,k}e^{2\pi ipj/N}e^{2\pi iqk/n}h_{j,k},\\
\end{eqnarray}
that is, it is the double discrete Fourier transform of the coefficients $h_{j,k}$, where the index $j$ labels a block or ring, and $k$ the element inside the $\mathrm{j^{th}}$ block.
For $H$ real and symmetric, its real eigenvalues $e(p,q)$ become $T$ doubly degenerate for $p=1,2,\dots,T$, and $t$ doubly degenerate for $q=1, 2, \dots, t$, and are of the form $e(p,q)=\sum_{j=0}^{N-1}\sum_{k=0}^{n-1}h_{p,q}\mathrm{cos}(2\pi jp/N)\mathrm{cos}(2\pi kq/n)$. The corresponding complex conjugated eigenvectors can be combined into two real eigenvectors.
With this representation of the eigenvectors, the expression for the diffusion $\sigma$ for a finite chain of rings, for an initial delocalized state is given by
\begin{equation}\label{sigmadeloc}
\sigma_{\mathrm{Deloc}}(t)=Dt\sqrt{2\sum_{j=1}^{T}j^2\big(\sum_{k=0}^{n-1}h_{j,k}\big)^2},
\end{equation}
where D is the spacing between the rings. For a localized state, the corresponding expression for the diffusion is
\begin{equation}\label{sigmaloc}
\sigma_{\mathrm{Loc}}(t)=Dt\sqrt{2\sum_{j=1}^{T}j^2\sum_{k=0}^{n-1}\big(h_{j,k}\big)^2}.
\end{equation}
These expressions are valid for small times or interaction strength, that is, $t\norm{H} \ll 1$. The proof is given in the appendix, for any initial condition.
A couple of simple cases will help us understand these expressions. 

In the ``far-field'' regime, with $D>R$, the couplings among the rings will be highly symmetric, similar to (\ref{SimHam}). For this case, ${\bf{b}_1}={\bf{b}_{N-1}}=VJ_{n}$, with the rest of the block matrices $\bf{b}_i=0$, where $J_{n}=\mathrm{circ}(1\, 1\, \dots\, 1)$ is an $n\times n$ matrix with ones as entries and $V=J (1/ D_{\mathrm{nm}})^3$ is the strength of the symmetric couplings between two nearest neighbouring rings, separated by a distance D. The expressions above for the diffusion will read $\sigma_{Deloc}=\sqrt{2}VDtn=\sqrt{2}J tn/D^2$, $\sigma_{Loc}=\sqrt{2}J t\sqrt{n}/D^2$, so that the symmetric couplings between the rings produce a linear scaling of the diffusion as a function of the number of sites $n$, and $\sqrt{n}$, for a delocalized and localized initial states, respectively. Thus, we proved that a linear dependence with the number of particles $n$ occurs under symmetric couplings, and will be use in the rest of this work as a proof of supertransfer. We also obtain that for a $1/r^3$ dipole-dipole decaying interaction, the diffusion decreases as $1/D^2$, with D the distance between rings, and is directly proportional to the typical interaction strength J. If all the interactions among the rings were included, the above expressions change to $\sigma_{Deloc}=\frac{\pi^4}{90}\sqrt{2}J tn/D^2$, $\sigma_{Loc}=\frac{\pi^4}{90}\sqrt{2}J t\sqrt{n}/D^2$.

In the ``near-field'' regime, with $D\le R$, the scaling with $n$ can increase beyond $n$ or $\sqrt{n}$ for $\sigma_{Deloc}$ and $\sigma_{Loc}$, respectively. The reason comes from how the interaction terms $\sum_{k=0}^{n-1}h_{j,k}$ and $\sum_{k=0}^{n-1}(h_{j,k})^2$ scale up with n. Without loss of generality, let's consider two nearest neighbours rings. The interaction term for the delocalized case is of the form $\sum_{k=0}^{n-1}h_{j,k} = \sum_{k=0}^{n-1}\big(D^2 + 2R^2(1-\cos(2\pi k/n))\big)^{-3/2}$, which exhibits a scaling with $n$ faster than linear from $D\le R$, for fixed $j$. This is a small size effect, and the scaling is linear for $n$ large enough (as can be readily checked by approximating the above sum by an integral for $n\gg 1$).

The expression for the diffusion of a delocalized state (\ref{sigmadeloc}) shows also some interesting ``interference effects'', in the sense that partial cancellations can occur when some of the coefficients $h_{j,k}$ are negative. In particular, when the condition $\sum_{k=0}^{n-1}h_{j,k}=0$ is met, the delocalized state does not propagate, while the localized does. Therefore, even in the closed system, an initial delocalized state does not necessarily diffuse longer than a localized state, during times $t\norm{H} \ll 1$.

Figure (\ref{Predictions}) makes a comparison between the simulated value of $\sigma(t)/Dt$, for delocalized and localized initial states, and for various values of the distance $D$ between the rings. In the far-field regime $D>R$, we clearly see a linear scaling of diffusion with $n$ for a delocalized initial state, and with $\sqrt{n}$ for a localized state. For the near-field regime $D<R$, there is a superlinear scaling which arises from the behaviour of the interaction terms in the Hamiltonian, as mentioned above. In all cases, there is a good agreement between the theoretical predictions and the numerical simulations. For times longer than $\epsilon/\norm{H}$, the excitons travel long enough so that the effects of the boundaries ``kick in'', and the block-circularity assumption worsens .

Note that the Hamiltonian corresponding to the ring where the exciton is initially created does not enter equations (\ref{sigmadeloc}) and (\ref{sigmaloc}), as $j=0$. Therefore, under the approximation $t\norm{H}< \epsilon$, these expressions for diffusion do not depend on the couplings of the ring where the initial state is located.

The above analysis, that was developed for a stacked ring geometry, can be used to approximate the diffusion in the helical rod setting, if the helices in the rod are approximated by stacked rings of the same radius $R$, with a distance between them equal to the pitch of the helix $d=D$, and with number of chromophores per ring $n$, equal to that in the rods per turn. Figure (\ref{RingsApprox}) exhibits this approximation. The blue circles correspond to numerical simulations for the spiral geometry with given $R$ and pitch $d$, while the red stars depicts the formulas (\ref{sigmadeloc}), (\ref{sigmaloc}) obtained from a set of facing rings with radius $R$ and separation $D=d$, indicating that the approximation given by the rings structure is quite good. In principle, if the number of chromophores per turn is not an integer, the angular position of the rings should be shifted by a given amount to account for this. However, due to the parametrization of the spiral that is being used, the nodes are perfectly aligned along the axis of the spiral, so no ``shift'' of the nodes position inside the rings is necessary.

\section{Open system dynamics: numerical simulations}
Having elucidated the behaviour of diffusion length in a closed system dynamics, in this section we analyse the behaviour of supertransfer subjected to an open quantum system dynamics, via two effects: on-site energy disorder and on-site dephasing.

\subsection{Effects of on-site disorder}
For small values of static energy disorder, the motion of the exciton should be hardly affected, propagating coherently. For times $t\gg \Sigma^{-1}$, diffusive behaviour sets in and the mismatch among the onsite energies causes localization \cite{MP}. This is depicted in Figure (\ref{DiffVsnVsDisor}). For disorder $\Sigma \ll 1/t$, the motion is coherent, and we obtain a supertransfer regime $\sigma\sim n$, while the exciton gets localized for $t\Sigma\gg 1$. A similar conclusion can be drawn from Figure (\ref{LambdaInDiffVsnVsDisor}), where the exponent $\lambda$ in $\sigma(t)\sim t^{\lambda}$ shows a coherent ($\lambda=1$) evolution for small values of disorder, and a crossover to classical diffusion takes place when $t\Sigma\sim 1/$, independently of $n$. In the classical regime, the diffusion fluctuates around $\lambda=0.5$.

Figure (\ref{LevelCurveDisor}) represents a level curve for diffusion as a function of $n$ and disorder $\Sigma$, obtained from Figure (\ref{DiffVsnVsDisor}). For disorder $\Sigma<10^{-2}$, the system is resilient to changes of disorder, and we see a linear enhancement of diffusion with $n$, while for $\Sigma>10^{-2}$, diffusion has a $\sqrt{n}$ behaviour. The areas between different values of disorder provides an indication of how robust diffusion is under changes of disorder. From this observation, it derives that diffusion seems more robust under changes of disorder in the quantum or ballistic regime (for $\Sigma<10^{-2}$) than in the ballistic regime.

\subsection{Effects of on-site dephasing}
On-site dephasing diminishes the coherence among the sites of the i
ring, which is fundamental for supertransfer effects to take place. 
For a tight-biding Hamiltonian on an infinite chain with periodic 
boundary conditions, nearest neighbour interaction $J$, and dephasing strength $\gamma$, see  \cite{GS}, the diffusion length is given by:

\begin{equation}
\sigma^2 = \frac{4J^2}{\hbar^2 \gamma}\Big[t + \frac{1}{\gamma}\big(1-e^{-\gamma t}\big)\Big].
\end{equation}
For small times, $\gamma t\ll 1$, the transport is initially ballistic, $\sigma(t)\sim t$, while for $\gamma t\gg 1$, the transport is diffusive, $\sigma(t)\sim \sqrt{t}$, with a crossover time of the order of $1/\gamma$.

A similar behaviour should be expected for the diffusion for the ring structure. For small times compared to this dephasing scale, $t<1/\gamma$, supertransfer effects will enhance the diffusion length by a factor of $n$. This is shown in Figure (\ref{svstg}), for $\gamma=1$, where at small times the motion is still coherent and $\sigma(t,n)\sim n$, while for  $t>1/\gamma$ the motion is classical and $\sigma(t,n)\sim \sqrt{n}$.

Using the block-circulant approximation employed in the closed system dynamics from the previous section, a similar calculation for diffusion can be carried out in the presence of dephasing as well. It is possible to verify that dephasing does not contribute to the diffusion, up to second order in the expansion of $\rho(t)=e^{t\mathcal{L}}\rho(0)$, with $\mathcal{L}=-i\mathrm{ad}_{\mathrm{H}}+\mathcal{L}_{\mathrm{deph}}$, for $\rho(0)$ being either a localized or delocalized state. That is, up to second order in time, the expressions (\ref{sigmadeloc}) and (\ref{sigmaloc}) remain unmodified under dephasing. Figure (\ref{PredVsDeph}) shows the diffusion as a function of the number of nodes $n$, for an initial delocalized state (left panels), localized state (right panels), in the near field $D/R=0.1$, $D/R=1$ and far field $D/R=10$. Most importantly, the theoretical predictions (\ref{sigmadeloc}) and (\ref{sigmaloc}) (blue squares), are contrasted with numerical simulations of diffusion including many values of dephasing: $\gamma=0.1$ (red triangles), $\gamma=1$ (green diamonds) and $\gamma=10$ (brown squares). In all cases, for small values of dephasing $\gamma=0.1$, the agreement between the numerical and analytical results is quite good, and the relative error between them is at most $3\%$ for all cases.

Figure (\ref{LambdaVsTime}) shows the best fit for the exponent $\lambda$ in $\sigma(t)\sim t^{\lambda}$, for a fixed value of dephasing and different number of sites n. The exciton start initially with a ballistic spreading $\lambda=1$, decaying to a diffusive classical walk regime $\lambda=0.5$. As the number of sites increases, we see a sharper drop of the exponent towards 1/2. The dependence with $n$ comes from the fact that the decoherence rate of a symmetric state coupled symmetrically to an common on-site dephasing bath goes as $n$ times the single exciton rate $\gamma$. We lack an explanation for the dip of the scaling exponent $\lambda$ below 0.5 around $t=1$, which might point to an intriguing physical feature, and is left for future research.

The scaling $\alpha$ in $\sigma(t)\sim n^\alpha$ as a function of time, for different values of dephasing is shown in Figure (\ref{AlphaVsTime}). The scaling goes from the supertransfer $\alpha=1$ to the classical walk regime $\alpha=1/2$ around the same time when the excitons diffusion starts to deviate from the ballistic spreading $\lambda=1$.

\section{Conclusions}
We have analysed both analytically and numerically how 
symmetry-enhanced supertransfer, for circular and helical 
geometry of chromophores, enhance excitonic diffusion lengths. 
For a closed system dynamics, we derived explicit expressions 
for diffusion for any initial state, demonstrating a 
factor $n$ enhancement in diffusion when the inter 
ring couplings are similar among each other. These formulas 
explicitly capture quantum interference effects for initial 
delocalized states, independently from the details of the 
rings where the excitons are initial created, and 
approximate the diffusion along helical rods quite well.

Moreover, we have studied the effects of environmental 
interactions on supertransfer, by including both energy disorder 
and interactions with a bosonic bath in a Haken-Strobl pure-dephasing 
model. The $n^2$ enhancement in diffusion prevails for times smaller 
than the disorder strength, and the system becomes more resilient 
against random perturbations in the onsite energies for increasing 
values of the number of nodes $n$ per ring or turn of the spiral. 
Due to the quasi 1D nature of the aggregates considered here,  
supertransfer effects are fragile to disorder on the site basis 
energies. We have shown analytically that the dephasing does 
not affect the form of our closed expressions for diffusion, 
up to second order in time, in good agreement with our simulations. 
A numerical analysis enabled us to obtain the scaling exponents of 
sigma with number of nodes $n$ and time $t$, revealing a crossover 
from the ballistic-supertransfer regime, to a diffusive-normal regime, 
in both cases around the same characteristic time $1/\gamma$. Our 
studies on exciton supertransfer dynamics presented here can be 
generalized for other complex quantum systems interacting 
with the non-Markovian and non-perturbative environments 
using the new techniques that have recently been developed 
in Refs. \cite{Shabani2011,Mohseni2011}.

\section{Acknowledgements}
The numerical simulations described in this paper were supported by the University of Southern California Center for High Performance Computing and Communications.  We acknowledge financial support from DARPA under the QuBE program and ENI (MM, SL). SL acknowledges support of NEC, INTEL, NSF, ONR, ISI, and the Santa Fe Institute. PZ acknowledges support from NSF grants PHY-803304, DMR-0804914 and PHY-0969969.

\appendix{}
\begin{widetext}

\section{Derivation of the closed expressions for the exciton diffusion length}
As shown in (\ref{eigenvalues}), the eigenvalues of a block circulant matrix are given by $e(p,q)=\sum_{j=0}^{N-1}\sum_{k=0}^{n-1}(w_j)^p (w_k)^q h_{j,k}=\sum_{j,k}\exp(2i\pi pj/N)\exp(2i\pi qk/n)h_{j,k}$. Assume the initial state is given by $\psi(0)=\sum_{r=1}^{N-1}\sum_{s=0}^{n-1}\alpha_{r,s}\ket{r}\ket{s}$, where for simplicity we assume $\alpha_{r,s}$ to be real. The second moment or diffusion $\sigma^2(t)$ is given by $\sigma^2(t)=D^2\sum_{R=0}^{N-1}R^2p_R(t)$, where $p_R(t)=\sum_{S=0}^{n-1}|\bra{R}\langle S|\psi(t)\rangle|^2$ is the probability that the exciton is in the ring $R$. Using the spectral decomposition of of the evolution operator $U(t)=\exp(-it H)$ we obtain $A_{R,S}(t)=\bra{R}\langle S|\psi(t)\rangle = \sum_{j=0}^{N-1}\sum_{k=0}^{n-1}\exp(-ite(j,k))\langle R | e(j)\rangle \langle S | e(k) \rangle \langle e(j,k)|\psi(0)\rangle$, where we have used that the eigenvectors have the tensor product structure $\ket{e(j,k)}=\ket{e(j)}\ket{e(k)}$. For $t\norm{H}\ll 1$, we employ the approximation $\exp(-ite(j,k))\sim 1 - ite(j,k)$. Putting all the terms together and expanding up to second order in time, we get $p_R(t)=\sum_{S=0}^{n-1}|A_{R,S}(t)|^2\sim \sum_{S=0}^{n-1}\Big[\alpha_{R,S}^2 + t^2\big(\sum_{j=0}^{N-1}\sum_{k=0}^{n-1}h_{j,k}\alpha_{R+j,S+k}\big)^2\Big]$, where we have used the convention that $\alpha_{-R,S}=\alpha_{R,-S}=\alpha_{R,S}$.

For an initial delocalized state on the 0 ring, $\alpha_{R,S}=\delta_{R,0}1/\sqrt{n}$, $p_R(t)\sim \sum_{S=0}^{n-1}\Big[\delta_{R,0}1/n + t^2\big(\sum_{j=0}^{N-1}\sum_{k=0}^{n-1}h_{j,k}\delta_{-R,j}1/\sqrt{n}\big)^2\Big] = \delta_{R,0} + t^2\big(\sum_{k=0}^{n-1}h_{-R,k}\big)^2$. Therefore, the diffusion is given by $\sigma(t)^2=D^2\sum_{R=-(T-1)}^{T-1}R^2\Big[\delta_{R,0} + t^2\big(\sum_{k=0}^{n-1}h_{R,k}\big)^2\Big]=D^2t^2\sum_{R=-(T-1)}^{T-1}R^2\big(\sum_{k=0}^{n-1}h_{R,k}\big)^2$.
For a localized state on the ring 0 and site 0, we have $\alpha_{R,S}=\delta_{R,0}\delta_{S,0}$, and $p_R(t)\sim \sum_{S=0}^{n-1}\Big[\delta_{R,0}1/n + t^2\big(\sum_{j=0}^{N-1}\sum_{k=0}^{n-1}h_{j,k}\delta_{-R,j}\delta_{-S,k}\big)^2\Big] = \sum_{S=0}^{n-1}\Big[\delta_{R,0}1/n + t^2\big(\sum_{j=0}^{N-1}\sum_{k=0}^{n-1}h_{j,k}\delta_{-R,j}\delta_{-S,k}\big)^2\Big] = \delta_{R,0} + t^2\sum_{S=0}^{n-1}h_{R,S}^2$, and the diffusion is given by  $\sigma(t)^2=D^2\sum_{R=-(T-1)}^{T-1}R^2\Big[\delta_{R,0} + t^2\sum_{S=0}^{n-1}h_{R,S}^2\Big]=D^2t^2\sum_{R=-(T-1)}^{T-1}R^2\sum_{S=0}^{n-1}h_{R,S}^2$.

\end{widetext}


\begin{thebibliography}{99}
\bibitem{Dicke54} Dicke, R. H. 1954 Coherence in Spontaneous Radiation Processes. {\it Phys. Rev.} {\bf 93}, 99-110 (doi:10.1103/PhysRev.93.99)

\bibitem{RehlerEberly71} Rehler, N. E. \& Eberly, J.H. 1971 Superrandiance. {\it Phys. Rev. A} {\bf 3}, 1735-1751 (doi:10.1103/PhysRevA.3.1735)

\bibitem{Fidder91} Fidder, H., Knoester, J. \& Wiersma, D.A. 1991 Optical properties of disordered molecular aggregates: A numerical study. {\it J. Chem. Phys.} {\bf 95}, 7880-7890 (doi:10.1063/1.461317)

\bibitem{Zhao99} Zhao, Y., Meier T., Zhang W. M., Chernyak  V. \& Mukamel, S. 1999 Superradiance Coherence Sizes in Single-Molecule Spectroscopy of LH2 Antenna Complexes. {\it J. Phys. Chem. B} {\bf 103}, 3954-3962 (doi:10.1021/jp990140z)


\bibitem{Palacios02} Palacios, M., de Weerd, F. L., Ihalainen J. A. , van Grondelle, R. \& van Amerongen, H. 2002 Superradiance and Exciton (De)localization in Light-Harvesting Complex II from Green Plants?. {\it  J. Phys. Chem. B} {\bf  106}, 5782-5787 (doi:10.1021/jp014078t)

\bibitem{Jin03} Jin, G., Zhang, P., Liu, Y. \& Sun, C. P. 2003 Superradiance of low-density Frenkel excitons in a crystal slab of three-level atoms: The quantum interference effect. {\it Phys. Rev. B} {\bf 68}, 134301 (doi:10.1103/PhysRevB.68.134301)

\bibitem{S} Strek, W. 1977 Cooperative energy transfer. {\it Phys. Lett. A} {\bf{62}}, 315-316 (doi:10.1016/0375-9601(77)90427-3)

\bibitem{Scholes02} Scholes, G. D. 2002 Designing light-harvesting antenna systems based on superradiant molecular aggregates. {\it Chem. Phys.} {\ bf 275},
373-386 (doi:10.1016/S0301-0104(01)00533-X)


\bibitem{LM} Lloyd, S. \& Mohseni, M. 2010 Symmetry-enhanced supertransfer of delocalized quantum states. {\it New J. Phys.} {\bf{12}}, 075020 (doi:10.1088/1367-2630/12/7/075020)


\bibitem{Jang04} Jang, S., Newton, M.D. \& Silbey, R.J. 2004 Multichromophoric F\"orster Resonance Energy Transfer. {\it Phys. Rev. Lett.} {\bf 92}, 218301 (doi:10.1103/PhysRevLett.92.218301)






\bibitem{Engel07} Engel, G. S., Calhoun, T. R., Read, E. L., Ahn, T. K., Mancal, T., Cheng, Y. C., Blankenship R. E. 2007 \& Fleming, G. R. Evidence for wavelike energy transfer through quantum coherence in photosynthetic systems. {\it Nature} {\bf 446}, 782-786 (doi:10.1038/nature05678)

\bibitem{Lee07} Lee, H., Cheng, Y. -C. \& Fleming, G. R. 2007 Coherence Dynamics in Photosynthesis: Protein Protection of Excitonic Coherence. {\it Science} {\bf  316}, 1462-1465 (doi: 10.1126/science.1142188)

\bibitem{Calhoun09} Calhoun, T. R., Ginsberg, N. S., Schlau-Cohen, G. S., Cheng, Y.-C., Ballottari, M., Bassi, R. \& Fleming, G. R. 2009 Quantum Coherence Enabled Determination of the Energy Landscape in Light-Harvesting Complex II. {\it J. Phys. Chem. B} {\bf 113}, 16291-16295 (doi: 10.1021/jp908300c) 

\bibitem{Mercer09} Mercer, I., El-Taha, Y., Kajumba, N., Marangos, J., Tisch, J., Gabrielsen, M., Cogdell, R., Springate, E. \& Turcu, E. 2009 Instantaneous Mapping of Coherently Coupled Electronic Transitions and Energy Transfers in a Photosynthetic Complex Using Angle-Resolved Coherent Optical Wave-Mixing. {\it Phys. Rev. Lett.} {\bf 102}, 057402 (doi:10.1103/PhysRevLett.102.057402)

\bibitem{Scholes09-1} Collini, E. \& Scholes, G. D. 2009 Coherent Intrachain Energy Migration in a Conjugated Polymer at Room Temperature. {\it Science} {\bf 323}, 369-373 (doi:10.1126/science.1164016)

\bibitem{Scholes09-2} Collini, E., Wong, C. Y., Wilk, K. E., Curmi, P. M., Brumer, P. \& Scholes, G. D. 2010 Coherently wired light-harvesting in photosynthetic marine algae at ambient temperature. {\it Nature} {\bf 463}, 644-647 (doi:10.1038/nature08811)

\bibitem{panit10} Panitchayangkoon, G., Hayes, D., Fransted, K. A., Caram, J. R., Harel, E., Wen, J., Blankenship, R. E. \& Engel, G. S. 2010 Long-lived quantum coherence in photosynthetic complexes at physiological temperature. {\it Proc. Nat. Acad. Sci.} {\bf 107}, 12766-12770 (doi:10.1073/pnas.1005484107)



\bibitem{mohseni-fmo} Mohseni,  M., Rebentrost, P., Lloyd, S. \& Aspuru-Guzik, A. 2008 Environment-assisted quantum walks in photosynthetic energy transfer. {\it J. Chem. Phys.} {\bf 129}, 174106 (doi: 10.1063/1.3002335)

\bibitem{Rebentrost08-2} Rebentrost, P., Mohseni, M., Kassal, I., Lloyd, S. \& Aspuru-Guzik, A. 2009 Environment-assisted quantum transport. {\it New J. of Phys.} {\bf 11}, 033003 (doi:10.1088/1367-2630/11/3/033003) 

\bibitem{Olaya-Castro08} Olaya-Castro, A., Lee, C. F., Olsen, F. F. \& Johnson, N. F. 2008 Efficiency of energy transfer in a light-harvesting system under quantum coherence. {\it Phys. Rev. B} {\bf 78}, 085115 (doi:10.1103/PhysRevB.78.085115)

\bibitem{Rebentrost08-1} Rebentrost, P., Mohseni, M. \& Aspuru-Guzik, A. 2009 Role of Quantum Coherence and Environmental Fluctuations in Chromophoric Energy Transport. {\it J. Phys. Chem. B} {\bf 113} 9942-9947 (doi:10.1021/jp901724d)


\bibitem{Plenio09} Caruso, F., Chin, A. W., Datta, A., Huelga, S. F. \& Plenio, M. B. 2009 Highly efficient energy excitation transfer in light-harvesting complexes: The fundamental role of noise-assisted transport. {\it J. of Chem. Phys.} {\bf 131}, 105106 (doi:10.1063/1.3223548)


\bibitem{AkiPNAS} Ishizaki, A. \& Fleming, G. R. 2009 Theoretical examination of quantum coherence in a photosynthetic system at physiological temperature. {\it Proc. Nat. Acad. Sci.} {\bf 106}, 17255-17260 (doi: 10.1073/pnas.0908989106)
(2009).

\bibitem{CaoSilbey} Cao, J. \& Silbey, R. 2009 Optimization of Exciton Trapping in Energy Transfer Processes. {\it J. Phys. Chem. A} {\bf 113}, 13825-13838 (doi: 10.1021/jp9032589)


\bibitem{Caruso10} Caruso, F., Chin, A. W., Datta, A., Huelga, S. F. \& Plenio, M. B. 2010 Entanglement and entangling power of the dynamics in light-harvesting complexes. {\it Phys. Rev. A} {\bf 81}, 062346 (doi: 10.1103/PhysRevA.81.062346)

\bibitem{Sarovar} Sarovar, M., Ishizaki, A., Fleming, G. R. \& Whaley, K. B. 2010 Quantum entanglement in photosynthetic light-harvesting complexes. {\it Nature} {\bf 6}, 462-467 (doi:10.1038/nphys1652) 

\bibitem{Fassioli10} Fassioli, F. \& Olaya-Castro, A. 2010 Distribution of entanglement in light-harvesting complexes and their quantum efficiency. {\it New J. Phys.} {\bf 12}, 085006 (doi:10.1088/1367-2630/12/8/085006)

\bibitem{Joel2010} Yuen-Zhou, J., Mohseni, M. \& Aspuru-Guzik, A. 2010 Quantum Process Tomography of Multichromophoric Systems via Ultrafast Spectroscopy. arXiv:1006.4866

\bibitem{Shabani2011} Shabani, A., Mohseni, M., Lloyd, S. \& Rabitz, H. 2011 Optimal and robust energy transfer in light-harvesting complexes: (I) Efficient simulation of excitonic dynamics in the non-perturbative and non-Markovian regimes. arXiv:1103.3823

\bibitem{Mohseni2011} Mohseni, M., Shabani, A., Rabitz, H. \& and Lloyd, S. 2011 Optimal and robust energy transport in light-harvesting complexes: (II) A quantum interplay of multichromophoric geometries and environmental interactions. arXiv:1104.4812 

\bibitem{Shim2011} Shim, S., Rebentrost, P., Valleau, S. \& Aspuru-Guzik, A. 2011 Microscopic origin of the long-lived quantum coherences in the Fenna-Matthew-Olson complex.  arXiv:1104.2943 

\bibitem{pemans03} Peumans, P., Yakimov, A. \& Forrest, S. R. 2003 Small molecular weight organic thin-film photodetectors and solar cells. {\it J. Appl. Phys.} {\bf 93}, 3693-3723 (doi:10.1063/1.1534621)

\bibitem{lunt09} Lunt, R. R., Giebink, N. C., Belak, A. A., Benziger, J. B. \& Forrest, S. R. 2009 Exciton diffusion lengths of organic semiconductor thin films measured by spectrally resolved photoluminescence quenching. {\it J. Appl. Phys.} {\bf 105}, 053711 (doi:10.1063/1.3079797)

\bibitem{high08} High, A. A., Novitskaya, E. E., Butov, L. V., Hanson, M. \& Gossard, A. C. 2008 Control of Exciton Fluxes in an Excitonic Integrated Circuit. {\it Science} {\bf 321}, 229-231 (doi:10.1126/science.1157845)

\bibitem{castellano10} Singh-Rachford, T. N. \& Castellano, F. N. 2010 Photon upconversion based on sensitized triplet-triplet annihilation. {\it Coordination Chemistry Reviews} {\bf 254}, 2560 - 2573 (doi:10.1016/j.ccr.2010.01.003)




\bibitem{E} Escalante, M., Lenferink, A., Zhao, Y., Tas, N., Huskens, J., Hunter, C. N., Subramaniam, V. \& Otto, C. 2010 Long-Range Energy Propagation in Nanometer Arrays of Light Harvesting Antenna Complexes. {\it Nano Lett.} {\bf{10}}, 1450-1457 (doi:10.1021/nl1003569) 


\bibitem{MPF} Miller, R. A., Presley, A. D. \& Francis, M. B. 2007 Self-Assembling Light-Harvesting Systems from Synthetically Modified Tobacco Mosaic Virus Coat Proteins. {\it J. Am. Chem. Soc.} {\bf 129}, 3104-3109 (doi:10.1021/ja063887t)

\bibitem{D} Davis, P. J. 1979 {\it{Circulant Matrices}}, 2nd edn, pp. 176-191. Chelsea Publishing Corporation 


\bibitem{T} Tee, G. J. 2007 Eigenvectors of Block Circulant and Alternating Circulant Matrices. {\it New Zealand J. Math.} {\bf 36}, 195-211

\bibitem{MP} Madhukar, A. \& Post, W. 1977 Exact Solution for the Diffusion of a Particle in a Medium with Site Diagonal and Off-Diagonal Dynamic Disorder. {\it Phys. Rev. Lett.} {\bf 39}, 1424 (doi:10.1103/PhysRevLett.39.1424)


\bibitem{GS} Grover, M. \& Silbey, R. 1971 Exciton Migration in Molecular Crystals. {\it J. Chem. Phys.} {\bf{34}}, 4843-4851 (doi:10.1063/1.1674761)

\end{thebibliography}
\end{document}